\documentclass[]{aa}

\def\Mv{$M_{\rm{v}}$}
\def\Msun{{\rm M}_{\odot}}
\def\kms{km~s$^{-1}$}
\def\ocen{$\omega$~Cen\/}
\def\log{{\rm{log}}}

\input psfig.tex

\begin{document}

\title{Accretion of gas by globular cluster stars}

\author{A. Thoul\inst{1}\thanks{Chercheur Qualifi\'e au Fonds National de la 
                         Recherche Scientifique (Belgium)} 
        \and
        A. Jorissen$^\star$\inst{2}
        \and
        S. Goriely$^\star$\inst{2}
        \and
        E. Jehin\inst{3}
        \and
        P. Magain\inst{1}
        \and
        A. Noels\inst{1}
        \and 
        G. Parmentier\inst{1}}
 
\offprints{A. Thoul, \email{thoul@astro.ulg.ac.be}}

\institute{Institut d'Astrophysique et de G\'eophysique, 
           Universit\'e de Li\`ege, 
           5 Avenue de Cointe, B-4000 Li\`ege, Belgium
      \and    
           Institut d'Astronomie et d'Astrophysique, CP 226,
           Universit\'e Libre de Bruxelles, 
           Boulevard du Triomphe, B-1050 Bruxelles, Belgium
      \and
           European Southern Observatory, Casilla 19001, Santiago 19, Chile}

\date{Received date; accepted date}

\abstract{
Some recent observations of the abundances of s-process, r-process, and
$\alpha$ elements in metal-poor stars have led to a new scenario
for their formation. According to this scenario, these stars were born
in a globular cluster and accreted the s-process enriched
gas expelled by cluster stars
of higher-mass, thereby modifying their surface
abundances. 
Later on, these polluted stars evaporated from the globular cluster
to constitute an important fraction of the current halo
population.
In addition, there are now many direct observations of 
abundance anomalies not only in globular cluster giant stars but also in
subgiant and main-sequence stars. Accretion provides again a plausible 
explanation for (at least some of) these peculiarities.
Here we investigate further the efficiency 
of the accretion scenario. We find that in 
concentrated clusters with large escape velocities, 
accretion is very efficient and can indeed lead to 
major modifications of the stellar surface abundances.
\keywords{globular clusters -- stars:abundances -- stars: chemically
  peculiar -- accretion -- stars: AGB and post-AGB}
}

\authorrunning{A. Thoul et al.}
\maketitle

\section{Introduction}

Strong correlations 
between the r-process and s-process 
element abundances and the $\alpha$-element abundances in field metal-poor
stars have been reported (\cite{jehin98}, 1999), 
separating these stars into two sub-populations.
The r-process elements correlate linearly with
the $\alpha$-elements, with a clumping at the maximum value of $[\alpha/Fe]$. 
The s-process elements, on the other hand, exhibit a more complex 
behavior when plotted against the $\alpha$ elements, and form
a ``two-branches diagram'', which we show schematically in Fig.~1.
We emphasize here that the observed stars being dwarf stars,
they cannot have synthetized these s-elements in their interior.

\begin{figure}[h]
\centerline{\psfig{figure=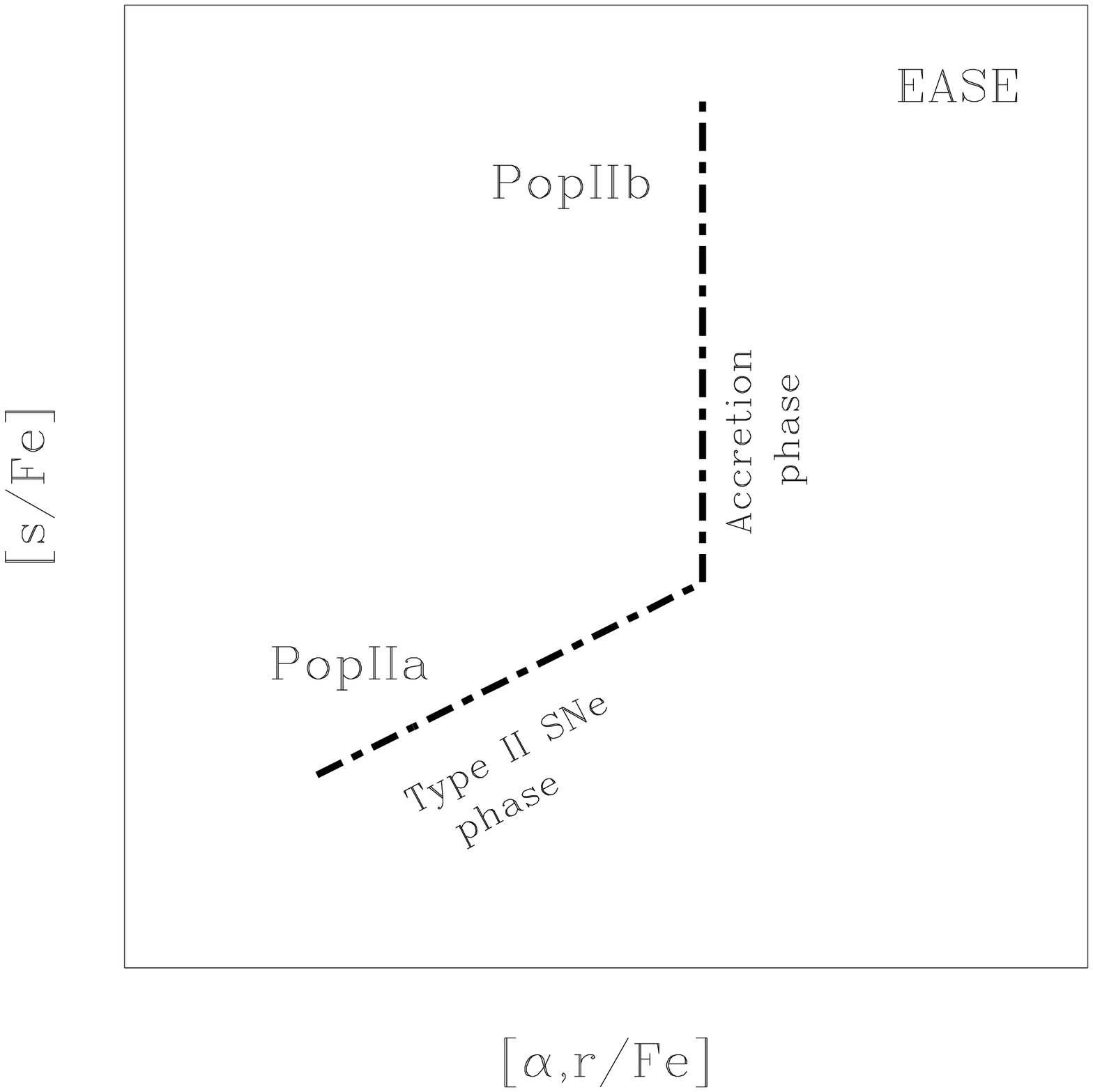,width=7.7cm}}
\caption[] {The ``two-branches diagram'' (see \cite{jehin99} for details).}
\label{figease}
\end{figure}

In order to explain these observations the EASE scenario has been developed
(\cite{jehin99}; \cite{parmen99}, \cite{parmen00a},
\cite{parmen00b})
which links the metal-poor field halo stars to the halo globular clusters 
(GCs).
According to this model, the field
halo stars are born in GCs. The evolution of the GCs is 
separated into two phases, directly connected to the two 
subpopulations now observed in the halo. During the first phase, 
``first generation'' stars are formed. The initial mass function in the very 
metal-poor medium favors the high masses, with $3\; \Msun < M <60\; \Msun$. 
The most massive of these stars quickly evolve to become supernovae, 
ejecting $\alpha$- and r-process 
elements into the surrounding intracluster medium. The explosions trigger
the formation of an outwardly expanding supershell. The primordial ISM gas 
is swept into the shell with the supernovae ejecta, and the resulting gas 
is enriched in $\alpha$- and r-process elements. 
A new burst of star formation occurs in 
the shell, where the density is very high. This triggered star formation
presumably has a Salpeter-like initial mass function. 
A proto-globular 
cluster is born. If the gravitational potential is too weak, it gets 
disrupted before having the chance of forming a GC, and the 
stars become field halo stars of the PopIIa, as defined on 
Fig.~\ref{figease}. The metallicity reached in the proto-GC depends on 
the number of supernovae which have exploded before the 
proto-GC gets disrupted.
The $\alpha$-element abundances are 
fixed by that time as well as by the initial mass function 
(supernova progenitors of different masses produce different relative 
amounts of $\alpha$ 
elements). The proto-globular cluster may in some cases survive this supernova 
phase. If the stars fall back under the effect of gravity, 
they form a GC. Detailed dynamical calculations 
have been performed (\cite{parmen99}) which show that
proto-globular clusters can indeed survive the supernova phase and 
metallicities such as those observed today in GCs can be achieved
in this scenario. It has also been shown that the EASE scenario, with its two 
separate star formation events, does not contradict observational
constraints such as the narrow metallicity range found in any given cluster.

During the second phase, intermediate-mass (second generation) stars evolve,
becoming asymptotic giant branch (AGB) stars. 
Those stars produce s-elements, which
are brought to the surface through the so-called third 
dredge-up occurring after the thermal pulses. 
The s-elements are then ejected into the surrounding intracluster medium 
by stellar winds. The intracluster gas
is therefore enriched
in s-elements, while the $\alpha$- and r-process element abundances
have been fixed at their highest value after all massive stars
have exploded. 
If the cluster can retain this enriched gas, it can be accreted by
the lower-mass stars which are still on the main 
sequence. If the accreted material does
not get mixed within the entire star but only in its convective zone, 
which contains about 1\%  
of the total stellar mass for 
stars still near the main-sequence turn-off of present GCs,
then even a small amount of accretion can 
lead to an appreciable enrichment of the stellar atmosphere. When the star 
evolves towards the giant stage, its convective zone becomes much larger, 
and the 
s-element enrichment is diluted considerably. 
We point out here that some mechanisms could be at work to already
trigger the
dilution of the accreted matter below the convective zone of 
{\it main-sequence}
stars (\cite{Proffitt89}; \cite{ProffittMichaud89}), thus diminishing the
impact on the surface abundances of the deep convective envelope 
developing in red giant stars. 
We come back to this issue in more detail in Sect.~2.
The accreting star can be 
ejected out of the GC at any time, by evaporation or by tidal 
forces possibly leading to the total disruption of the cluster
while crossing the disk.
There is indeed now observational evidence that GCs lose stars:
two cannon-ball stars in 47~Tuc (\cite{meylan91}; \cite{jorissen01}; see also
\cite{capaccioli93}; \cite{piotto97}, and references therein), and, above all,
the recent observations of ``tidal tails'' around {\bf many} GCs
(\cite{grillmair95}; \cite{leon00}; \cite{odenkirchen01}). 
Stars enhanced in s-process elements by accretion and then 
stripped from GCs form the Pop IIb halo.

The hypothesis that the gas ejected by the intermediate-mass stars in GCs
can be accreted by the other (lower-mass) cluster stars has been 
examined before. There is strong observational and theoretical evidence 
that stars reaching the AGB phase lose a large amount of mass.
Several studies have been devoted to the fate of this gas in clusters 
(\cite{scott75}; \cite{faulkner77}; \cite{vandenberg77}; \cite{vandenberg78};
 \cite{scott78}; \cite{faulkner84}; \cite{faulknercol84}; \cite{smith96}). 
Stellar ejecta in GCs with shallow potential wells can leave 
the cluster via a smooth wind-like outflow. Faulkner \& Freeman (1977)
and VandenBerg \& Faulkner (1977)
find steady-state time-independent flow solutions
in clusters of $10^5\,\Msun$. However, they show that in GCs
with deep enough potential wells, the gas can accumulate into the cluster,
forming a central reservoir with a radius comparable to the GC
core radius. Faulkner \& Coleman (1984)
show that a small number of 
low-velocity low-mass stars in the cluster core can in this case accrete 
enough matter to form $10\;\Msun$ black holes, which can be considered
as an extreme case. All those studies looked at what 
happens in present-day GCs. 
Smith (1996) addressed the 
question whether such winds might have been possible within young GCs,
during epochs of much higher stellar mass-loss rates. 
Many GCs have high enough escape velocities
to retain at least some of the stellar ejecta. A substantial amount of
intracluster material could have been acquired when the turn-off
mass was about $5\,\Msun$. In less tightly-bound clusters, the stellar
ejecta are lost from the cluster either stochastically or through a
continuous wind. 
In the cases where the stellar ejecta are retained in the cluster,
the gas could be accreted 
by other cluster 
stars, thereby modifying their surface composition. 
Several models estimating the 
efficiency of the accretion scenario
have appeared in the literature (see e.g. \cite{dantona83}; \cite{smith96}).
A quantitative model investigating the recycling of nova ejecta has
been presented by Smith \& Kraft (1996).
Qualitative discussions of the accretion model have been presented by
Bell et al. (1981), Norris \& Da Costa (1995), and Cannon et al. (1998).

In this paper,
we study the process of accretion onto
GC stars from a central reservoir of gas. In particular
we take into account the evolution of this process with 
time. Indeed the rate of mass ejection into the cluster is strongly 
time-dependent, as it is highest when intermediate-mass stars reach the 
mass-losing stage, decreasing very rapidly as the turn-off mass decreases.
In Sect.~2, we examine the possible links between halo stars and GCs. We
address the problem of the ``missing'' intracluster gas in Sect.~3. We
present some observational signatures that may possibly be
related to
accretion by cluster stars in Sect.~4.
In Sect.~5 we present the method used to calculate the efficiency
of accretion of gas
by GC stars. The results are summarized in Sect.~6. Finally Sect.~7
contains the conclusions.

\section{Unease with the EASE scenario}

Although the EASE scenario nicely explains some abundance correlations
in halo stars through a sequence of events in GCs, it faces  
several problems.

The GC evaporation is not the only way to populate 
the halo. In particular, the discovery that the Sagittarius dwarf
galaxy is in the process of being tidally disrupted by our Galaxy and its stars
incorporated 
in the halo has led to a serious reconsideration of the role of mergers in the
formation of the halo. Initially put forward by Searle \& Zinn (1978),
in opposition to the classical rapid collapse envisaged by Eggen et al. (1962),
the merger scenario has gradually gained consensus. 
Relics of these past mergers
may be the streamers highlighted by Majewski et al. (1996)
or the blue metal-poor stars highlighted by Preston et al. (1994).

If field halo stars come from globulars, they should exhibit the
same pattern of abundance
peculiarities as GC stars [like C-(Na,Al) anticorrelation, 
N-Na-Al correlation]. 
\mbox{}Hanson et al. (1998)
have shown that, based on their location in the
([Na/Fe], [Mg/Fe]) diagram, halo field and GC stars of
the same metallicity are not surrogates of one another. 
In particular, no field halo stars seem to exhibit the large [Na/Fe]
overabundances observed in some bright giants in GCs (\cite{Pilachowski96}). 
Nevertheless, the scatter in [Na/Fe] at a given metallicity observed among
halo stars (Fig.~4 of \cite{Hanson98}) is certainly large enough to
accomodate the 0.3 to 0.4~dex range observed in 47~Tuc main-sequence
stars and attributed to the wind pollution scenario.

If the accretion scenario is at work in GCs, one
would expect that the pollution levels
decrease from main sequence to giant stars as the convective
envelope deepens, and dilutes the material initially restricted to a
thin surface layer. Several among the chemical anomalies observed in
GCs and mentioned above do not exhibit this trend,
however, as they remain unchanged all the
way from the main sequence to the giant branch. This suggests that the 
accreted matter is rapidly diluted even in main sequence stars.
At least two processes may be invoked to
trigger the dilution of the accreted matter into the deep radiative layers 
of main sequence stars. A turbulent diffusion mechanism triggered by
the inversion of mean molecular weight due to the higher He content of 
the accreted matter has been presented by Proffitt (1989)
and Proffitt \& Michaud (1989)
in relation with Am and Ba 
stars. These authors show that even a modest composition inversion
(say $\Delta X = -0.001$, and correspondingly $\Delta Y = +0.001$,
where $X$ and $Y$ are the hydrogen and helium mass fractions,
respectively), over the outer $10^{-3}\;\Msun$ of the star results in
dilution of the added layer by a factor of 50 in $10^8$~y. The depth
reached by such a mixing may be sufficient in order to prevent further 
dilution during the first dredge-up. This effect may explain why 
the level of abundance anomalies is similar in dwarf Ba
stars and in giant Ba stars (\cite{North94}), and why the level of 
abundance anomalies does not change between dwarf and giant stars in 
GCs.
Another cause of mixing specific to GCs is tidal mixing 
caused by occasional
close encounters of stars in clusters as suggested by Cox (1998).

The present paper is motivated by the   
connection between s-process enhancements in field halo stars and
in cluster stars that  
evaporated from the cluster. There is not much evidence, however, 
for s-process
anomalies due to accretion in globular cluster stars, except for
\ocen\, (but see the discussions of Sects.~4.1 and \ref{Sect:results}) and for 
the metal-poor cluster M15 where 
Sneden et al. (1997)
reported bimodal abundance distributions for both [Ba/Fe] and
[Eu/Fe] (with a strong correlation between Ba and Eu) in the bright
giants.

\section{Low level of intracluster gas}
\label{Sect:absence}

The absence of interstellar material in GCs has been a
long-standing puzzle (e.g. \cite{roberts88}). \cite{Freire01}
recently achieved a positive detection of 
0.1~$\Msun$ of ionized gas in the core 
of 47~Tuc. But this mass is much less than the $\sim 100\;\Msun$ of
intracluster material expected to accumulate between  two successive
passages of the GC through the galactic disk (\cite{roberts88}).
Knapp et al. (1995, 1996) have recently reinvestigated this
question by setting upper limits on
both the dust and ionized gas contents for a number of  GCs using IRAS
and radio observations, respectively.
Knapp et al. (1996) conclude that their `upper limits on the
mass of ionized gas are, for all observed clusters, much smaller than
the amount expected to have accumulated from stellar mass loss if all
that gas remains in the cluster between galactic plane passages'.
Simple models of the equilibrium velocity 
distribution of the gas supposedly photoionized
by post-AGB stars indicate, however, that this gas is able to escape
from the cluster in the time separating two passages through the
galactic plane. Taking this effect into account,
the amount of gas predicted to accumulate from stellar mass loss in this 
time is now well below the observational upper limits. 
As far as the dust is concerned, Knapp et al. (1995) conclude that the
upper limits are much lower than the amounts expected to have
accumulated over 10$^8$~y from stellar mass loss. 

Recycling of the mass lost by the evolved stars, by the accretion
scenario investigated in the
present paper, offers another way (see e.g., \cite{Freire01} for a
list of some of the other 
possible mechanisms) to account for the apparent lack of
intracluster matter.

\section{Abundance anomalies observed in globular cluster stars, and
  their possible relevance to the accretion scenario}

\subsection{Peculiar red stars in \ocen} 

The GC \ocen\, is  discussed separately
as it bears several peculiarities.
It is the most massive and largest GC (\cite{harris96}) and 
exhibits a large spread in metallicity ($-1.8 < $  [Fe/H]  $ < -0.8$;
\cite{NorrisDaCosta95}) with some spatial asymmetry
(\cite{Jurcsik98}), but  
the most significant peculiarity of \ocen\, in the present context is
undoubtedly the presence of CH, Ba and S stars (\cite{Dickens76}; 
\cite{LloydEvans77}, 1983, 
1986; \cite{Vanture94}), as  listed in Table~\ref{Tab:binaryoCen}.
Mild overabundances of the heavy elements produced by the s-process of 
nucleosynthesis are observed in the peculiar red stars (PRS) of \ocen\,
(Table~\ref{Tab:binaryoCen}). With 
the exception of the CH star ROA~279, PRS in \ocen\,  
form the tail of the general trend of increasing heavy-element abundances with 
increasing [Fe/H] typical of \ocen\, 
(\cite{Vanture94}; \cite{Smith95}; \cite{NorrisDaCosta95}). 
PRS in \ocen\, should thus not be considered as
exceptional, but rather as the most extreme members of a
chemically-inhomogeneous population.
The PRS in $\omega$~Cen generally turn out to be somewhat enriched in
carbon, since the
\ocen\, giants studied by Norris \& DaCosta (1995) typically have
[C/Fe] $\sim -0.6$, as compared to [C/Fe] values as large as +0.2 for
some of the PRS listed in Table~\ref{Tab:binaryoCen}. The existence of 
such carbon-enriched giants is another specific feature of \ocen,
since in GCs with [Fe/H] $\sim -1.0$, red giants within
1 mag of the red giant tip exhibit carbon depletions by factors in the 
range 0.3--1.0 dex instead (\cite{Briley92}, and references therein). 

The PRS of \ocen\, are fainter than the
luminosity threshold of the thermally-pulsing AGB (\Mv\, $\sim -2.5$), 
according to the $M = 1.25\,\Msun, Z = 0.001$ evolutionary track of
Charbonnel et al. (1996) [their Fig. 2]. 
Therefore,
internal nucleosynthesis on the TP-AGB cannot account for their chemical
peculiarities. 

Barium and CH stars in the field have been shown to belong
systematically to binary systems (\cite{McClure90}), with
their chemical anomalies
resulting from the accretion of carbon- and s-process-rich matter
from their AGB companion that has now evolved into a white dwarf. 
A radial-velocity monitoring of the PRS in \ocen\, has been performed
by Mayor et al. (1998), and has revealed that 
with the exception of the classical CH stars ROA~55 and
ROA~70, as well as possibly the S stars ROA 320 and V6, PRS
in \ocen\, do NOT belong to binary systems, however
(\cite{Mayor96}; Table~\ref{Tab:binaryoCen}). 
A scenario specific to GCs seems thus at work to
produce these PRS. One possible scenario is that those PRS were
formerly member of binary systems that were disrupted by a close
encounter with another cluster star. \cite{Mayor96} estimate that
systems with periods in excess of 3 to 10~y, depending on their
position in the cluster, have been disrupted by tidal encounters over
the lifetime of the cluster. Before being disrupted, Ba, S or 
CH systems with such long periods may
indeed have been produced by the same mass-transfer scenario as that 
invoked to account for the PRS observed in the field 
(\cite{Jorissen98}, 1999).

Norris \& DaCosta (1995) and Cannon et al. (1998) have suggested
instead that the
PRS envelope has been polluted by winds expelled in the
intracluster medium by an early
generation of AGB stars. However, the simultaneous increase of Fe and
the s-process elements observed in \ocen\,
requires as well a source of Fe that cannot be
delivered by AGB stars, a
statement not easy to reconcile with the pollution scenario, as discussed by
Smith et al. (1995). 
Such peculiarities in \ocen\, are more in favor of a merging between two 
different clusters, as further discussed in Sect.~\ref{Sect:results},
which also shows that accretion is not at all efficient in \ocen.

\begin{table*}
\caption[]{\label{Tab:binaryoCen}
Abundances, luminosities, and binary properties of PRS in $\omega$ Cen.
Column 1 provides the ROA star number from Woolley (1966; last three
digits as in Woolley's Tables I and 
II, or variable numbers from Woolley's Table III). Column 2 gives
the spectral type from the reference mentioned in column 3, columns 4
and 5 list \Mv\, (adopting a
distance modulus of 13.92; \cite{harris96}) and $B-V$ from \cite{Mayor98},   
columns 6, 7 and 8 (taken from \cite{Mayor98}) give the uncertainty 
$\epsilon_{V_r}$ of the mean
radial velocity, the rms value $\sigma_{V_r}$ for stars with more than
one measurement and the probability $P(\chi^2)$ that the observed     
$\sigma_{V_r}$ is due to observational error only. Column 9, labelled
$V_r$,
provides a flag characterizing the binary nature of the star,
according to the following rules: a star
is considered as spectroscopic binary (`SB')  
if $P(\chi^2) < 0.01$ and $\sigma_{V_r} > 2.0$~\kms. The latter
condition is imposed because red giants close to the RGB tip are known 
to exhibit a radial-velocity jitter of the order of 1~\kms\
(\cite{Mayor84}; 
see also Fig.~1 of Jorissen et al. 1998). If  
$0.01 \le P(\chi^2) < 0.05$, or $P(\chi^2) < 0.01$ and 
$\sigma_{V_r} > 1.5$~\kms, the star is considered as variable (`var').
Columns 10, 11 and 12 list the [Fe/H], [C/Fe] and  
s-process abundances when available, 
with [s/Fe] =  ([Y/Fe] + [Zr/Fe])/2. Column 13
provides the reference for the abundance data.}

\begin{tabular}{lclllllllllllllll}
\hline
ROA & Sp. & Ref & \Mv & $B-V$ & $\epsilon_{V_r}$ & $\sigma_{V_r}$ & $P(\chi^2)$
& $V_r$ & [Fe/H] & [C/Fe] & [s/Fe] & Ref \cr
\hline
55 & CH & 1 & -2.43 & 1.74 & 0.74 & 3.98 & 0.000 & SB & \cr
70 & CH & 1 & -2.34 & 1.82 & 0.90 & 4.33 & 0.000 & SB & \cr
V53& Ba & 2 & -1.97 & 1.78 & 0.50 & 1.00 & 0.015 & var& \cr
201& S2.5/1-2 & 3 & -1.82 & 1.61 & 0.52 & 1.04 & 0.030 & var& -0.85 &
-0.30 & 0.33 & 4\cr
219& Ba & 2 & -1.76 & 1.68 & 0.41 & 1.01 & 0.003 & var& -1.25 & 0.22 & 
0.40 & 4\cr
231& Ba & 2 & -1.73 & 1.63 & 0.40 & 1.56 & 0.000 & var& -1.10 &-0.60 & 
0.44 & 4\cr
243& Ba & 2 & -1.70 & 1.31 & 0.31 & 0.70 & 0.155 &    & \cr
248& Ba & 2 & -1.69 & 1.84 & 0.46 & 1.44 & 0.000 & var& -0.78 & -    & 
0.17 & 4\cr
270& Ba & 2 & -1.61 & 1.53 & 0.37 & 0.99 & 0.014 & var& -1.24 & -0.01& 
0.34 & 4\cr
276& Ba & 2 & -1.61 & 1.52 & 0.37 & 0.82 & 0.113 &    & \cr
279& CH & 1 & -1.60 & 1.48 & 0.33 & 0.88 & 0.209 &    & -1.70 & -    & 
0.48 & 4\cr
286& Ba & 2 & -1.59 & 1.57 & 0.25 & 0.45 & 0.774 &    & \cr
287& Ba & 2 & -1.59 & 1.48 & 0.30 & 0.67 & 0.335 &    & -1.43 & 0.06 & 
0.26 & 4\cr
300& S2/1-2 & 3 & -1.56 & 1.58 & 0.27 & 0.53 & 0.548 &    & \cr
316& Ba & 2 & -1.50 & 1.66 & 0.58 & 1.16 & 0.017 & var& \cr
320& S2/1-2 & 3 & -1.49 & 1.65 & 0.86 & 2.58 & 0.000 & SB? & \cr
321& Ba & 2 & -1.48 & 1.45 & 0.23 & 0.27 & 0.945 &    & \cr
V6 & S3/1 & 3 & -1.47 & 1.54 & 1.18 & 2.35 & 0.000 & SB? & \cr 
324& Ba & 2 & -1.46 & 1.59 & 0.25 & 0.39 & 0.749 &    & \cr
332& Ba & 2 & -1.45 & 1.53 & 0.31 & 0.75 & 0.115 &    & \cr
357& Ba & 2 & -1.42 & 1.48 & 0.48 & 1.17 & 0.000 & var& -0.85 & - &
0.20 & 4\cr
371& K5Ba & 3 & -1.38 & 1.74 & 0.51 & 1.25 & 0.000 & var& -0.79 &
-0.46 & 0.46 & 4\cr
   &      &   &       &      &      &      &       &    & -1.0  & 
-     & 0.8 & 5\cr
421& Ba & 2 & -1.25 & 1.50 & 0.35 & 0.86 & 0.035 & var& \cr
425& S3/2 & 3 & -1.24 & 1.69 & 0.33 & 0.31 & 0.890 &    & \cr
447& S2/2 & 3 & -1.18 & 1.71 & 0.28 & 0.63 & 0.338 &    & \cr
451& Ba & 2 & -1.18 & 1.53 & 0.41 & 0.92 & 0.046 & var& \cr
V17& Ba & 2 & -1.17 & 1.80 & 0.65 & 1.46 & 0.000 & var& \cr
480& Ba & 2 & -1.15 & 1.51 & 0.49 & 0.84 & 0.133 &    & -0.95 & -0.65
& 0.38 & 4\cr
505& Ba & 2 & -1.09 & 1.28 & 0.27 & 0.61 & 0.363 &    & \cr
513& S3/2 & 3 & -1.08 & 1.67 & 0.53 & 0.92 & 0.065 &    & \cr
577& CH &   1 & -0.97 & 1.55 & 0.53 & 1.30 & 0.008 & var& \cr
\hline
\end{tabular}

References to spectral types and abundances:\\
(1) Mayor et al. (1998), and references therein; (2) Lloyd Evans (1986);
(3) Lloyd Evans (1983); (4) Norris \& Da Costa (1995); (5) Vanture et 
al. (1994)
\end{table*}

\subsection{Abundance anomalies in main sequence or subgiant stars}
    
Abundance anomalies have long been known in GC
giant stars (see e.g. Kraft (1994) and DaCosta (1997) for recent
reviews). However, the
dredge-up processes occurring in these 
stars make it very difficult to disentangle intrinsic from extrinsic
causes to the observed chemical peculiarities. 
Such dredge-up processes are not expected to occur in main sequence
stars, according to canonical stellar evolution theories. Abundance anomalies
in those stars therefore point in principle to external
causes. Because of the
faintness of main sequence stars in GCs, abundance
analyses of these stars  have but only recently become feasible.
Indications for anomalies have since then rapidly accumulated. They
may be grouped into two classes -- CN and Na anomalies -- which are
discussed in turn.

\subsubsection{CN anomalies in main sequence stars}

The bimodal nature of the distribution of the CN band strength of red
giant stars in several GCs has been recognized since the 
early 1980s (see references in \cite{Kraft94} and \cite{Cannon98}).
This property has recently been extended to main sequence stars in
47~Tuc (\cite{Bell83}; \cite{Briley94}; \cite{Cannon98}), M5 and NGC~6752 
(\cite{Suntzeff91}), and M13 and M71 (\cite{Briley01}). 
The ratio of CN-weak 
to CN-rich stars is similar among dwarf and more luminous
stars, implying  little change in the overall distribution of CN with
evolutionary state (\cite{Briley94}). CN-strong stars tend to be more
numerous in the central region of 47~Tuc (\cite{Norris79};
\cite{Freeman85}).

Recently, Grundhal et al. (1998) have suggested that the significant
scatter in the Str\"omgren $c_1$ index observed along the red giant
branch down to the turnoff for all clusters more metal-poor than
[Fe/H] = -1.2 is attributable to star-to-star abundance variations in
the CNO elements through the strength of the CH G band and the violet
CN bands. 

The anticorrelation between the CN and CH band strengths is equally found
in dwarf and giant stars (e.g., \cite{Cannon98}, 
\cite{Briley01} and references therein), 
and indicates that a C deficiency goes along with a N
overabundance. Contrary to internal mixing scenarios, 
the primordial scenarios generally
invoked to account for these anomalies extending down to the main
sequence are difficult to reconcile with C depletions. This 
difficulty may be circumvented 
if `C-poor' stars represent in fact the primordial C abundance level while
C-rich stars have {\it accreted} carbon, as  
already suggested in the case of \ocen\, by Norris \& DaCosta (1995).    
The CN-CH anticorrelation should thus be re-interpreted in terms of
accretion of either C-poor N-rich matter, or C-rich N-poor matter. 
In the context of the accretion scenario, this situation may result
depending on whether or not hot-bottom-burning
(e.g. \cite{Boothroyd93})  was operating in the
polluting AGB star, as already considered in the simple model developed 
by Norris \& DaCosta (1995).  This interpretation of the 
CN-CH anticorrelation requires that ejecta from different AGB stars 
(being either C-poor N-rich, or C-rich N-poor) do mot mix in the
intracluster gas, a constraint difficult to reconcile however with the 
model of the central gas reservoir developed in Sect.~\ref{Sect:accretion}.

\subsubsection{Na anomalies in main sequence and subgiant stars}
\label{Sect:Na}

Enhancements of Na and Al are common in GC giants, and   
are correlated with N enhancements, and C and O depletions
(\cite{Cottrell81}, \cite{Norris81} and 
\cite{DaCosta97}, and references therein). The first indication that
Na enhancements are observed as well in main sequence stars has been
provided by Briley et al. (1995) for 47~Tuc. The reported enhancements
([Na/Fe] $\sim 0.3$ to 0.4~dex) in
three CN-strong main sequence stars (\Mv $\sim
+4$) are similar to those observed in evolved stars (\Mv $\sim
-1.5$) of 47~Tuc
(\cite{Cottrell81}; \cite{Gratton86}; \cite{Brown92}).

Very recent observations of turn-off and subgiant stars in 
GCs show that they exhibit clear O-Na and Mg-Al 
anti-correlations (\cite{gratton01}),
a behavior which had previously been seen only in GC giant stars.
This result cannot be explained by deep mixing scenarios and requires
some other mechanism, such as accretion of polluted intracluster gas
(\cite{ventura01}).

These anomalies that clearly point against an internal mixing
scenario do not seem restricted to massive, metal-rich clusters like
47~Tuc, since
subgiant stars observed by King (1998) in the metal-poor cluster M92
([Fe/H] = -2.5) also reveal large Na overabundances (0.8~dex) 
correlated with Mg deficiencies.
Interestingly enough, these Na overabundances are accompanied by modest Ba 
overabundances (0.4~dex).

There are nevertheless indications that 
qualitatively different processes may be at work in metal-poor and
metal-rich clusters. For example, 
the constancy of the ratio of CN-weak/CN-rich 
stars from the main sequence to the RGB observed in 47~Tuc, along with 
the constancy of the [Na/Fe]
abundances, contrast with the progressive Na enhancements and C
depletions with increasing luminosity reported for the more metal-poor clusters
M13 (\cite{Pilachowski96}), M4 and NGC 6752 (\cite{Suntzeff91}).
To complicate the picture further, the opposite trend is observed in
M92, since the Na overabundances in the
subgiants (\cite{King98}) are larger than in the RGB-tip giants
(\cite{Shetrone96}). 
It should be remarked that, in the context of the accretion
scenario, a more relevant distinctive
parameter might be the cluster mass (47~Tuc is indeed very massive) 
rather than its metallicity.

In addition, 
Armosky et al. (1994) has
observed that in some clusters, the abundances of Y, Ba, Ce,
and Nd in giant stars are nearly constant as Na/Fe varies widely.

The different observed behaviors in the Na abundances clearly reflects
the diversity of the nucleosynthetic  processes which produce that
chemical element.
Sodium and s-elements are efficiently produced through proton mixing by 
the NeNa cycle (see for example figures 1 and 7 in \cite{goriely00}). 
On the other hand, the Na-production during the third dredge-up 
is not correlated to the s-process (Mowlavi 1999).
Finally, Na is produced in massive stars through the combustion of carbon
(\cite{woosley95}).

\subsection{Period variations of RR Lyrae in globular clusters}

Cox (1998) has suggested that the slight period decrease (2.5
$10^{-5}$~d over 80~y) observed
for the RR Lyr V53 in M15, which cannot be attributed to secular
evolutionary effects, is caused by the accretion of about
10$^{-7}\,\Msun$ over $\approx$ 100~y. The corresponding accretion
rate (10$^{-9}\,\Msun$~y$^{-1}$) is however much larger than that
predicted to result from the accretion of gas from a central reservoir 
in M15. Table~\ref{Tab:results} predicts that a 1~$\Msun$ star accretes 
0.07~$\Msun$ over the cluster lifetime of $\sim 10^{10}$~y, 
resulting in an {\it average} 
accretion rate of $7\;10^{-12}\,\Msun$~y$^{-1}$.

\section{The accretion scenario}
\label{Sect:accretion}

It is worthwhile to attempt to understand the fate of all the
interstellar matter ejected by massive and intermediate-mass stars during
the early phases of a GC's evolution.
On the one hand, it can bring an answer to the apparent lack of ISM in
present-day GCs and, on the other hand, if the efficiency of the
accretion process can be related to some properties of GCs, it
may be expected to unravel
why GCs exhibit different abundance patterns in their main-sequence, 
turn-off and subgiant stars.

\subsection{Bondi accretion rate}

Following Bondi's (\cite{bondi52}) 
model for spherical accretion, the rate 
of mass accretion by a star of mass $M_s$ is given by
\begin{equation}
{{\rm d}M_s\over {\rm d}t}
= 4\pi\lambda(GM_s)^2\rho_g(v_{\rm rel}^2+c_s^2)^{-3/2}
\label{msdot}
\end{equation}
where $\lambda$ is a constant of order unity, $\rho_g$ is the
unperturbed gas density, $c_s$ is the sound speed in the gas, and
$v_{\rm rel}$ is the relative velocity of the star with respect to the gas.

In what follows, all masses are given in units of $1\;\Msun$.

\subsection{Mass loss by AGB stars}

\begin{figure}[h]
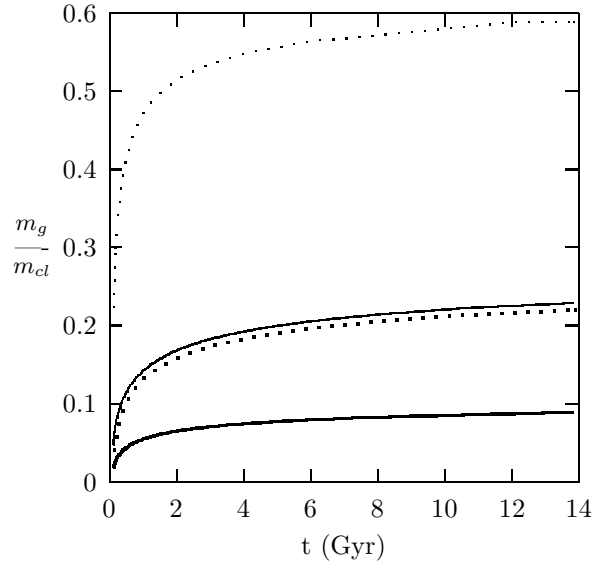

\input Mg.tex
\caption{Mass of gas ejected by the AGB stars, normalized to the cluster's
initial mass. We have used
           $\alpha=2.35$, $m_l=0.1$,  and $m_u=8$ (thin solid line);
           $\alpha=1.5$,  $m_l=0.1$,  and $m_u=8$ (thin dotted line);
           $\alpha=2.35$, $m_l=0.01$, and $m_u=8$ (thick solid line);
           $\alpha=2.35$, $m_l=0.1$,  and $m_u=7$ (thick dotted line).}
\label{figmg}
\end{figure}

First we look at the amount of gas ejected by the AGB stars. We assume 
that the stars eject their mass instantaneously when reaching the AGB phase.
The rate of gas production by AGB stars is given by 
\begin{equation}
{{\rm d}m_g\over {\rm d}t}=-m_{\rm ej} {{\rm d}N\over {\rm d}t}
\label{Mg}
\end{equation}
where $m_{\rm ej}$ is the mass ejected by AGB stars, and ${\rm d}N/{\rm d}t$ is 
the rate at 
which stars reach the AGB phase. We can write 
\begin{equation}
{{\rm d}N\over {\rm d}t}={{\rm d}N\over {\rm d}m}{{\rm d}m\over {\rm d}t}
\label{dndt}
\end{equation}
where ${\rm d}N/{\rm d}m$ is the initial mass function (IMF), 
giving the number $N$ of stars 
born in a given mass interval ($m$,$m+{\rm d}m$), and ${\rm d}m/{\rm d}t$ 
is the rate at which 
stars in this mass range reach the AGB phase. We choose a simple power-law 
initial mass function:
\begin{equation}
{{\rm d}N\over {\rm d}m}=K_1\, m_{\rm cl,0}\, m^{-\alpha}
\label{imf}
\end{equation}
where $m_{\rm cl,0}$ is the total initial cluster mass, $\alpha$ is the power-law 
index, and the normalization constant 
\begin{equation}
K_1=(\alpha-2)(m_l^{2-\alpha}-m_u^{2-\alpha})^{-1}
\label{k1}
\end{equation}
where $m_l$ and $m_u$ are the lower and upper mass limits of the mass spectrum.
For a Salpeter-like IMF, we have $\alpha=2.35$, whereas if we want
to favor the higher mass stars we choose a flatter spectrum with, e.g.,
$\alpha=1.5$. The evolution time $t_{\rm MS}$ (in y) of a star of mass $m$ 
(between 1 and $10\,\Msun$) along the main sequence can be approximated by 
(\cite{bahcall83})
\begin{equation}
\log\, t_{\rm MS}=10-3.6\, \log\, m + (\log\, m)^2.
\label{tms}
\end{equation}
Here we neglect the additional post-main-sequence
evolution time, which is 
small compared to $t_{\rm MS}$. From this relation we get
\begin{equation}
{{\rm d}m\over {\rm d}t_{\rm MS}}={m\over t_{\rm MS}}{1\over 2\, \log\, m-3.6}.
\label{tms2}
\end{equation}
Finally, we adopt the updated results of Weidemann (2000)
for the initial-to-final mass relation.
We fit their results:
\begin{equation}
m_{f}=\cases{0.053\;m_i + 0.497 & if $1.0 \le m_i\le 2.5$;\cr
             0.113\;m_i + 0.347 & if $2.5 \le m_i\le 4.0$;\cr
             0.073\;m_i + 0.507 & if $4.0 \le m_i\le 8.0$;\cr}
\label{mej}
\end{equation}
where
$m_f$ and $m_i$ are the final and initial masses of the star, so that
$m_{\rm ej}=m_i-m_f$. 
Putting together 
equations~(\ref{Mg})-(\ref{mej}), 
we can calculate the rate at which the gas 
is ejected by the AGB stars as a function of time. The total amount of gas 
which has been ejected by the AGB stars in the cluster after a given time $t$ 
is obtained by integrating the gas production rate:
\begin{equation}
m_g(t)=\int_{t_1}^t {{\rm d}m_g\over {\rm d}t'}{\rm d}t',
\label{mg}
\end{equation}
where $t_1$ is the time at which the most massive stars (of mass $m_u$)
reach the AGB, i.e., 
$\log\, t_1=10.-3.6\, \log\, m_u + (\log\, m_u)^2$.
In Figure~\ref{figmg}, we show (thin solid line) $m_g(t)$ for 
$m_u=8$, $m_l=0.1$, and $\alpha=2.35$. 
To illustrate the sensitivity on the IMF parameters, 
we also show the results obtained for other values of these parameters.
Increasing the upper limit of the mass spectrum of course increases
the number of high mass stars, resulting in more matter being ejected. 
Lowering the lower mass limit $m_l$ increases the total mass situated in
low-mass stars, thus decreasing the relative number of high-mass stars.
By far the most sensitive parameter is $\alpha$, the power-law index of 
the IMF.
Choosing a flatter mass spectrum ($\alpha=1.5$) considerably increases
the number of high-mass stars in the cluster, leading to a much larger
amount of gas ejected into the cluster's ISM. With this value of $\alpha$
we find that about $60\%$ of the cluster's initial mass can be returned
to the cluster as gas in $10^{10}$y, 
whereas for a Salpeter's IMF, this number
is $20\%$. We also see that the rate of gas production in the cluster is very 
high at early times, and decreases very quickly as the most massive stars 
leave the AGB phase. Note that in this figure $m_g$ is the total amount of 
matter which has been lost by the cluster's AGB stars. This is different from 
the amount of gas available for accretion, since some of it could have already
been accreted, or could escape before being accreted.

If we assume that all the gas ejected by cluster stars remains in the cluster
and that the only gas-removal mechanism is by tidal sweeping when the GC
crosses the galactic disk, we get a strict upper limit on the amount 
of gas which can be found at any time in the cluster. 
Even though there could have been a very large amount of 
gas in the cluster before its first passage through the galactic disk, 
this amount becomes much smaller afterwards. 

\subsection{Accretion by cluster stars}

The rate of mass accretion by a star of mass $M_s$ is given by 
equation~(\ref{msdot}), where the gas density 
$\rho_g$ is determined by the amount 
of gas ejected by the AGB stars, but also by the ability of the cluster 
to retain this gas, and by the amount of gas which is accreted.
We can write this equation as
\begin{equation}
{{\rm d}m_s\over {\rm d}\tau}=2.376\; 10^{-7}\, \lambda\,  m_s^2\, \rho\, v_{10}^{-3}
\label{mdotless}
\end{equation}
where $m_s=M_s/\Msun$, $\tau=t/10^6$y, 
$v_{10}=(v_{\rm rel}^2+c_s^2)^{1/2}/$(10~km/s)
and $\rho=\rho_g/(1\Msun/{\rm pc}^3)$. 

The ``average'' gas density in the cluster can be written as
\begin{equation}
\bar{\rho}={(m_g-m_a-m_{\rm esc})\over 4\pi r_{g}^3/3}
\label{rhog}
\end{equation}
where the radius $r_{g}$ of the gas reservoir is given in pc, $m_a$ is the 
total mass of accreted gas, $m_{\rm esc}$ is the mass of gas 
escaping from the cluster, either through a continuous wind or by tidal forces 
(for example when the cluster crosses the galactic disk).
The actual gas density is of course a function of $r$,
and we expect it to adopt some smooth profile peaking at the cluster's
center and decreasing rapidly. 
Here we assume that the gas sinks to the cluster center, forming a
central reservoir of homogeneous density and of radius comparable to the 
core radius $r_c$. 
Typical values for $r_c$ range from 0.1 to 10\, pc. 
If we assume that all the cluster stars have stochastic 
orbits, the fraction of stars in the cluster core is
also the fraction of time spent by each star in the cluster core.
On the other hand, stars that have sunk to the center could have smaller 
velocities and would tend to remain in the core for longer periods 
of time, accreting most of the gas (\cite{faulknercol84}). Here, we
use the first hypothesis, keeping in mind that some stars
finding themselves preferably in the core could accrete more mass than
calculated here, while others would accrete less.
Using the model of King (1962), the ratio $\gamma$ of the number of stars in
the core to the total number of stars in the cluster ranges
between 0.3 and 0.1 approximately, for concentration parameters 
$c=\log(r_t/r_c)$ between 1 and 2 ($r_t$ is the tidal radius).  
We ignore mass segregation effects.

At a given time, a star is accreting gas at a rate
which depends on its position in the cluster, its velocity 
with respect to the gas, and its own mass (which itself can vary with
time as the star accretes or ejects gas). To get the total amount of
gas which has been accreted by a given star of mass $m_s$ at a 
given time $\tau$, we must integrate the mass accretion rate over time:
\begin{equation}
m_{a,s}(\tau)=\gamma\int_{\tau_1}^\tau {{\rm d}m_s\over {\rm d}{\tau}'} 
               {\rm d}{\tau}'
\label{mas}
\end{equation}
where $\tau_1=t_1/10^6$y.
We note that in equation~(\ref{mdotless}), at any time $\tau$ we have 
$m_s(\tau)=m_{s,0}+m_{a,s}(\tau)$ where $m_{s,0}$ is the star's initial mass.
Finally, the total mass of gas accreted by the stars is obtained
by integrating $m_{a,s}$ over the mass spectrum:
\begin{equation}
m_a(\tau)=\int_{m_l}^{m_{TO}} m_{a,s} {{\rm d}N\over {\rm d}m} {\rm d}m.
\label{ma}
\end{equation}
Here, for the sake of simplicity, 
we have made the assumption that stars do not accrete matter after
reaching the turn-off point. Indeed, those stars will instead 
{\it eject} matter through stellar winds.
In equation~(\ref{ma}), $m_{a,s}$ 
and the turn-off mass $m_{TO}$ both depend on time. This is also the case 
for the mass spectrum ${\rm d}N/{\rm d}m$. 
Indeed, higher-mass stars die leaving remnants
of lower masses, and some stars, especially the lower-mass stars,
escape from the cluster by evaporation or through external tidal forces.
The accretion process can also modify the mass spectrum if some stars accrete
an appreciable fraction of their own initial mass. 
For simplicity, we assume that the mass spectrum 
remains constant with time. Since 
most of the gas is ejected at early times, as shown in Fig~(\ref{figmg}),
most of the accretion also takes place at those early times and we expect 
that during the time interval of highest accretion rate, the variation in 
the mass spectrum is not significant.

\section{Results}
\label{Sect:results}

\begin{table*}
\caption[] {Structural parameters for some globular clusters. 
         Unless otherwise noted, the masses $M$ and the velocity dispersions
         $\sigma_c$ are from Mandushev et al. (1991), 
         the central densities $\rho_c$ are from Webbink (1985),
         the core ($r_c$) and tidal ($r_t$) radii are 
         from the catalogue of Harris (1996),
         the cluster's orbital periods ($P$) and eccentricities ($e$) are 
         from Dinescu et al. (1999). 
         The escape velocities $v_{\rm{esc}}$ are from Madore (1980).
         and $c=log(r_t/r_c)$ is the concentration parameter 
         (cc stands for ``core collapse''). } 
\begin{tabular}[t]{|l|l|l|l|l|l|l|l|l|l|l|l|l|l|l|l|l|l|l|}
\hline
NGC             &$M$    &$\log \rho_c$    &$r_c$  &$r_t$  &$c$    
        &$\sigma_c$ &$v_{\rm{esc}}$      &$P$    &$e$\\
         &($10^5\Msun$)  &($\Msun$/pc$^3$)  &(pc)     &(pc)
         &       &(km/s)   &(km/s)   &($10^6$ y) & \\
\hline
104 (47Tuc)     &11$^a$  &3.52$^d$      &0.2$^a$&48.     &2.40     &10.0   &56.8   &190-193 &0.17\\
5139 (\ocen) &51$^b$  &3.35             &3.7    &64.     &1.24     &16.6   &51.2   &120-123 &0.67\\
5272 (M3)       &6.8     &3.86          &1.4    &99.8    &1.85     &4.8    &34.3   &297-321 &0.43\\
5904 (M5)       &4.6     &4.11          &0.90$^e$&56.9$^e$&1.80     &6.5    &34.9   &722-995 &0.88\\
6205 (M13)      &5.8     &3.56          &1.3$^e$&47.1$^e$&1.56     &7.86   &30.4   &429-526 &0.62\\
6341 (M92)      &3.9     &4.38          &0.54   &34.9    &1.81     &6.1    &33.1   &201-208 &0.77\\
6752            &1.4     &4.47          &0.19   &60.     &2.5(cc)  &4.9    &31.1   &153-156 &0.08\\
7078 (M15)      &4.9$^c$ &6.62$^d$      &0.64$^e$&78.7$^e$&2.1(cc) &15.1   &40.9   &242-253 &0.31\\
\hline
\end{tabular}

         $^a$ Murphy et al. 1998 
         $^b$ Meylan et al. 1995 
         $^c$ Dull et al. 1997
         $^d$ Gebhardt \& Fischer 1995 
         $^e$ Lehman \& Scholz 1997.
\end{table*}

\begin{table*}
\caption[]{\label{Tab:results}
Results for the clusters listed in Table~2, at 
$t=14\, $Gy: $m_{a}$,
total amount of gas that has 
been accreted by the cluster stars, 
$m_{\rm esc}$, amount of gas which has been lost from the cluster 
(swept when crossing the galactic plane), $m_{\rm res}$, 
the amount of residual gas in the cluster (varies between 0 just after 
crossing the galactic plane and its
maximum value just before crossing the plane), $m_{\rm knapp}$ the upper limit 
of the observed dust and ionized gas (\cite{knapp95}, 1996), 
and the fraction of mass accreted by a $1\;\Msun$ star (last column).}
\begin{tabular}[t]{|l|l|l|l|l|l|l|l|l|l|l|l|l|l|l|l|l|l|l|}
\hline
NGC             &$m_{a}/m_{\rm cl,0}$       &$m_{\rm esc}/m_{\rm cl,0}$
&$m_{a}/m_g$    &$m_{\rm res}$      &$m_{\rm knapp}$    &$m_{a}/m_i$, ($m_i=1$~M$_\odot$)\\
                &(\%)                     &(\%)                     &(\%)             &($\Msun$)        &($\Msun$)              &(\%)       \\
\hline
104 (47Tuc)     &21                     &0.2                    &98             &$<$7           &1                      &80     \\
5139 ($\omega$Cen)&0.8                  &21                     &0.4            &$<$542                 &14                     &0.1    \\
5272 (M3)       &5                      &16                     &24             &$<$145         &14                     &11     \\
5904 (M5)       &14                     &8                      &63             &$<$140         &7                      &40     \\
6205 (M13)      &4                      &17                     &18             &$<$233         &7                      &8      \\
6341 (M92)      &14                     &8                      &63             &$<$33          &34                     &37     \\
6752            &20                     &1                      &95             &$<$1           &5                      &76     \\
7078 (M15)      &4                      &18                     &16             &$<$69          &298                    &7      \\
\hline
\end{tabular}
\end{table*}

\begin{figure}[h]
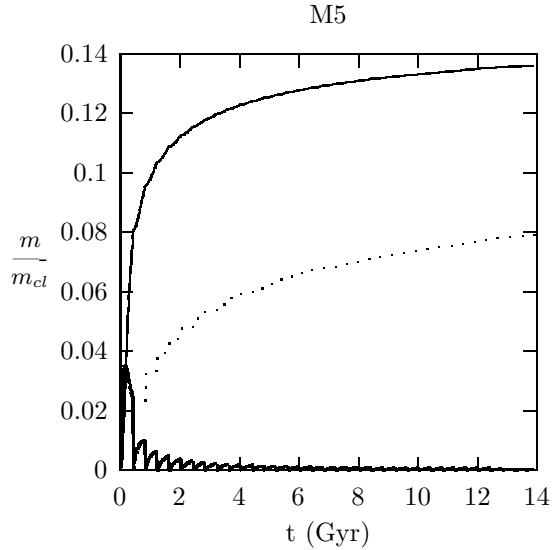

\input M5.tex
\caption[]{Results obtained for the cluster M5: $m_{a}$ (thin solid line),
$m_{\rm esc}$ (dotted line) and $m_{res}$ (thick solid line) as a function of 
time.}
\label{figM5}
\end{figure}

We solve the system of equations~(1)-(13). 
We assume that the gas forms a homogeneous
central reservoir of radius $r_g=r_c$ and that its 
temperature $T\approx\, 5000$\,K 
(\cite{vandenberg77}), which gives $c_s\approx 5.8$\,km/s. 
We also assume that the stars have 
chaotic motions so that they  spend about 20\% of their time in the 
cluster core, independently of their mass and average velocity. 
The amount of mass accretion is proportional to the 
inverse cube of $v_{10}$ and $r_c$, so that slight variations in 
these parameters 
induce large differences  in the accreted mass. 
For the relative stellar velocities, 
we use the ``average'' value, given by the 
velocity dispersion in the cluster core, $\sigma_c$. 
We use the present values for the
structural and dynamical parameters of the GCs, as 
listed in Table~2, even though 
those parameters could have changed as the clusters evolved chemically and
dynamically.
These clusters have been selected on the basis of their high escape velocities
(these clusters are tightly bound and therefore more likely to retain
gas) and/or because abundance anomalies have been observed for stars belonging
to these clusters.
The reservoir of gas is emptied each time the GC crosses the 
galactic disk, i.e. every $P/2$ where $P$ is the cluster's orbital period.
This is the only contribution considered here for the term $m_{\rm esc}$ 
appearing in Eq.~\ref{rhog}. Other removal 
processes, like type I supernovae, kinetic effects of strong winds
from millisecond pulsars (\cite{Spergel91, Freire01} or a photoionized 
cluster wind driven by the UV radiation from hot stars 
(see Sect.~\ref{Sect:absence}
and \cite{knapp96, Freire01}), were not included in $m_{\rm esc}$. Any such
process would decrease the efficiency of the present accretion
scenario. The results presented here must therefore be considered as
upper bounds.

Finally, we use $\alpha=2.35$, $m_l=0.1$, $m_u=8$, and $\lambda=1$
in all cases. The results at $t=14$ Gy are listed in Table~3, where we give
the total amount of gas which has 
been accreted by the stars, $m_{a}$, 
the amount of gas lost from the cluster (swept when crossing
the galactic plane), $m_{esc}$, 
the maximum amount of residual gas in the cluster, $m_{res}$,
and the fraction of mass accreted by a $1\;\Msun$ star.

Over the lifetime of the cluster, the total amount of matter 
injected into the GC-interstellar medium by the AGB stars is about 20\% of 
the cluster's initial mass (see Fig.~2).
In the most concentrated clusters, most of this 
gas is accreted by the cluster's lower-mass stars. 
Indeed, in 47Tuc, NGC6752, M5, and M92, more than 60\% of the gas ejected by 
the AGB stars is accreted by the cluster stars. In those clusters, 
$1\;\Msun$ stars
can accrete an appreciable fraction of their initial mass. The envelopes
of those stars will reflect the composition of the intracluster
medium rather than the composition of their interior, even if a physical
mechanism is at work to induce mixing with the deeper layers of the star.
We note here that these clusters are rich in observed
stellar composition anomalies (see Sect.~4).
The only cluster for which we get a very low amount of accretion
is \ocen. This is due to a combination of factors: \ocen\, has a 
low concentration parameter compared to the other clusters we
have selected. It also has a rather large core radius, and a high
core velocity dispersion, as well as frequent disk crossings. 
On the other hand, \ocen\, is a 
peculiar GC in many other respects and should probably not be
considered as a test bed for the accretion scenario, as many studies
now view \ocen\ as a conglomerate of different subsystems or even as
the nucleus of a dwarf galaxy 
(e.g., \cite{Lee99, hilker00, majewski00, Pancino00}).

The upper limits we get for the current residual intracluster gas are 
usually larger
than the observational limits determined by Knapp et al. (1995, 1996),
but of course the exact amount of gas 
predicted in the cluster will depend on the time elapsed since its
last passage through the galactic plane. Moreover, other 
gas sweeping  mechanisms that were not included in the present model 
are possibly at work, as discussed above. 

The results for $m_a$, $m_{esc}$ and $m_{res}$
as a function of time are shown in Fig.~\ref{figM5} for one 
particular cluster, M5. 
As expected, we find that most of the accretion takes place at early times,
when the amount of gas ejected by the AGB stars is the largest. 
The amount of residual intracluster gas increases with time 
in between each passage 
through the galactic plane, but becomes very small at late times.

\section{Conclusions}

It was  shown in this paper that accretion by low-mass GC stars
of the gas ejected in the intracluster medium by moderately massive stars
may be quite efficient. If enough mass is accreted, it can lead
to major alteration of the stellar surface composition. This 
supports the EASE scenario and provides a plausible explanation for the
lack of intracluster
gas, and for some of the abundance anomalies observed in GC stars.

\begin{acknowledgements}
This work has been supported by the P\^ole d'Attraction Interuniversitaire
P4/05 (SSTC, Belgium) and by FRFC F6/15-OL-F63 (FNRS, Belgium).
\end{acknowledgements}

\end{document}